\documentclass[reprint,superscriptaddress,amsmath,amssymb,aps,prc,twocolumn,floatfix]{revtex4-2}

\usepackage{makecell}
\usepackage{graphicx}
\usepackage{hyperref}
\usepackage{color}
\usepackage[english]{babel}
\usepackage{bm}
\usepackage[T1]{fontenc}
\usepackage{multirow}
\usepackage[output-decimal-marker={.},exponent-product=\cdot]{siunitx}

\begin{document}

\title{Comment on ``Spin-trap isomers in deformed, odd-odd nuclei in the light rare-earth region near $N=98$''}

\author{M. Stryjczyk}
\email{marek.m.stryjczyk@jyu.fi}
\affiliation{University of Jyvaskyla, Department of Physics, Accelerator laboratory, P.O. Box 35(YFL) FI-40014 University of Jyvaskyla, Finland}
\author{A.~Jaries}
\email{arthur.a.jaries@jyu.fi}
\affiliation{University of Jyvaskyla, Department of Physics, Accelerator laboratory, P.O. Box 35(YFL) FI-40014 University of Jyvaskyla, Finland}
\author{A.~Kankainen}
\email{anu.kankainen@jyu.fi}
\affiliation{University of Jyvaskyla, Department of Physics, Accelerator laboratory, P.O. Box 35(YFL) FI-40014 University of Jyvaskyla, Finland}
\author{T.~Eronen}
\email{tommi.eronen@jyu.fi}
\affiliation{University of Jyvaskyla, Department of Physics, Accelerator laboratory, P.O. Box 35(YFL) FI-40014 University of Jyvaskyla, Finland}

\begin{abstract}
A new isomeric $(4^-)$ state at 285.5(32) keV in $^{162}$Tb was reported by R. Orford \textit{et al.} [Phys. Rev. C 102, 011303(R) (2020)] based on a Penning-trap mass measurement. Here we show that this result is not compatible with existing experimental data. The state identified as $^{162}$Tb$^{m}$ with a mass-excess value of $-65593.9(25)$~keV is actually the $1^-$ ground state. The state identified as the ground state of $^{162}$Tb is most likely a molecular contaminant with the same mass-over-charge ratio.
\end{abstract}

\maketitle

\begin{figure}
     \centering
      \includegraphics[width=\columnwidth]{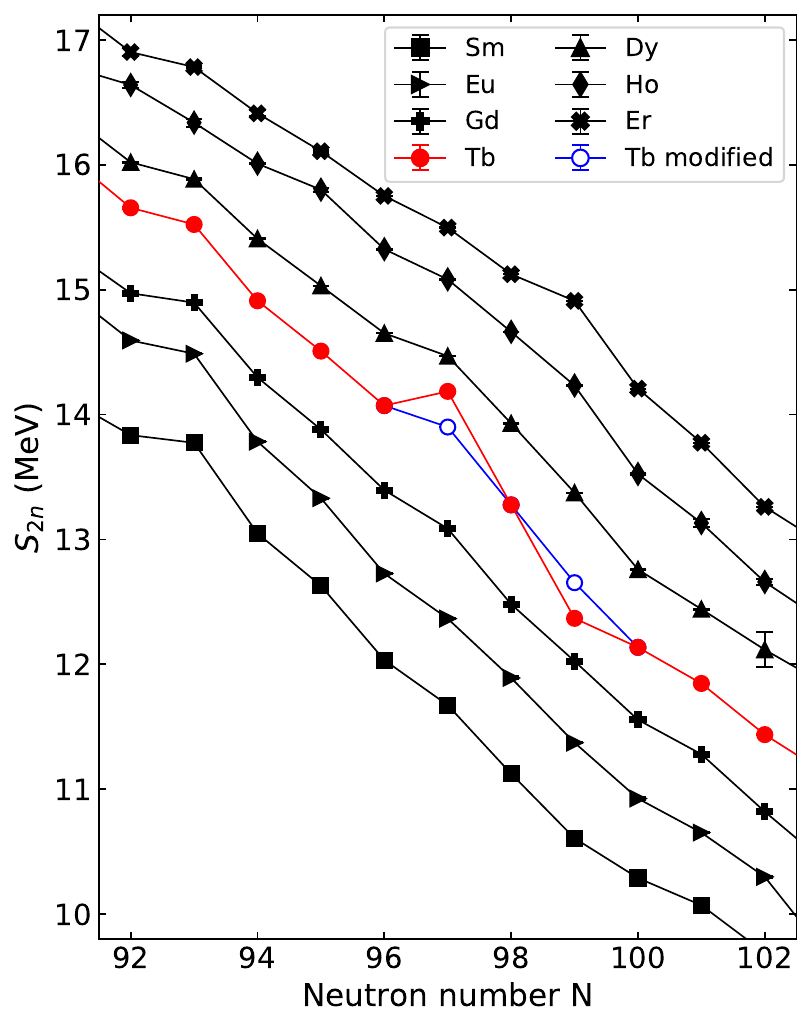}
     \caption{\label{fig:S2n}The $S_{2n}$ curves for Tb ($Z=65$, in red) and neighboring isotopic chains, from $Z=62$ to $Z=68$, based on the values from AME20 \cite{AME20}. The open blue circles show the $S_{2n}$ curve calculated using the $^{162}$Tb isomeric-state mass-excess value reported in Ref. \cite{Orford2020}.}
\end{figure}

A previously unknown spin-trap isomer in $^{162}$Tb at 285.5(32) keV was reported for the first time in Ref. \cite{Orford2020}. The authors tentatively assigned spin-parity $1^-$ to the ground state and $4^-$ to this newly observed isomer. This result was the first direct mass determination of the $^{162}$Tb ground state and in the most recent Atomic Mass Evaluation 2020 (AME20) it superseded previously used decay data \cite{AME20,Huang2021}. However, the incompatibility of the reported excitation energy with the results from $\beta$-decay \cite{Funke1966,Schima1966,Gujrathi1967,Kaffrell1968,Chang1970,Kawade1977,Gehrke1982} and transfer-reaction \cite{Burke2007} studies was not discussed. In addition, the reported mass values result in unusual fluctuations of the two-neutron separation energy ($S_{2n}$) curve around the neutron number $N=98$ for the Tb isotopic chain, not observed in the neighboring isotopic chains, see Fig.~\ref{fig:S2n}. In this comment we show that the presence of the isomer is incompatible with the existing spectroscopic data.

The spin-parity assignments of the ground state (${J^\pi = 1^-}$) and the first excited state at 39 keV ($J^\pi = 2^-$) in $^{162}$Tb are firmly established \cite{Nica2024} from decay and transfer-reactions studies \cite{Chang1970,Gehrke1982,Burke2007}. The $Q$ value for the  $^{163}$Dy$(t,\alpha)^{162}$Tb$(1^-)$ reaction, which was not included in AME20 \cite{Huang2021}, is 8073(3) keV ~\cite{Burke2007}. Combined with the well-known mass of stable $^{163}$Dy (ME$_\text{lit.} = -66380.9(7)$~keV \cite{AME20}), which is based on a direct mass measurement at the TRIGA-TRAP Penning trap \cite{Schneider2015} as well as electron- and neutron-capture $Q$ values \cite{Huang2021}, it yields a mass-excess value of $-65596.9(31)$~keV for the $1^-$ state in $^{162}$Tb. This is in an excellent agreement with the mass-excess value reported for the new $(4^-)$ isomer in Orford \textit{et al.} \cite{Orford2020}, $-65593.9(25)$~keV. Based on this observation we can deduce that if there are two long-living states present in $^{162}$Tb, then the isomer has to be $1^-$. This makes the state assignment proposed in Ref. \cite{Orford2020} incompatible with the existing experimental data. 

The most likely explanation for the lower-lying state reported in Ref.~\cite{Orford2020} is a molecular contaminant with the same mass-over-charge ($m/q$) ratio. The frequency ratio given for the double-charged $^{162}$Tb$^{gs}$ in Ref.~\cite{Orford2020} corresponds to $\text{ME} = -32939.7(10)$~keV for a singly-charged $A=81$ ion. This is very close, within 2.1(10)~keV, to the mass-excess of a singly-charged $^{1}\text{H}^{32}\text{S}^{16}\text{O}_{3}$ molecule, $-32937.5724(16)$~keV \cite{AME20}. 

When the isomer from Ref.~\cite{Orford2020} is assumed to be the real ground state of $^{162}$Tb, the discrepancies between the Penning-trap data and the decay data reported in AME20~\cite{Huang2021} decrease significantly. In particular, the difference between the $\beta$-decay study reported in Ref.~\cite{Schima1966} and the Penning-trap data \cite{Orford2020} is reduced from 4.4 to 1.3 standard deviations \cite{Huang2021}, resulting in a disappearance of the inconsistency between two pieces of well-documented data \cite{Huang2021}. The same hypothesis leads to an increased $\beta$-decay $Q$ value of $^{162}$Tb, from $2301.8(21)$~keV to $2587.3(26)$~keV. This solves the discrepancy in the $\beta$-decay feeding to the 2371-keV state in the daughter $^{162}$Dy \cite{Nica2024}, energetically impossible with the older $Q_{\beta}$ value. Finally, we note that this assumption also flattens the $S_{2n}$ curve at $N=97$ and $N=99$, as shown with the blue curve in Fig.~\ref{fig:S2n}. 

Based on the presented reasoning, we propose that the mass of the state assigned as $^{162}$Tb$^{m}$ in Ref.~\cite{Orford2020} is for the $1^-$ ground state while the mass of $^{162}$Tb$^{gs}$ corresponds to the $^{1}\text{H}^{32}\text{S}^{16}\text{O}_{3}^+$ molecular contamination.

\begin{acknowledgments}

This project has received funding from the Research Council of Finland projects No. 295207, 306980, 327629, and 354968.

\end{acknowledgments}

\bibliography{bibfile}

\end{document}